# Effects of Edge Passivations on the Electronic and Magnetic Properties of Zigzag Boron-Nitride Nanoribbons with Even and Odd-Line Stone-Wales (5–7 Pair) Defects

**SRKC Sharma Yamijala**[†] **and Swapan K Pati**[‡,§,*]

[†]Chemistry and Physics of Materials Unit, [‡]Theoretical Sciences Unit, [§]New Chemistry Unit, Jawaharlal Nehru Centre for Advanced Scientific Research, Jakkur P.O., Bangalore−560064, India. E-mail-ID: pati@jncasr.ac.in

**Abstract:**

First-principles spin-polarized calculations have been performed on passivated Boron-Nitride Nanoribbons (BNNRs) with pentagon-heptagon line-defects (PHLDs) (also called as Stone-Wales line-defects). Two kinds of PHLDs, namely, even-line and odd-line PHLDs, have been added either at one edge or at both edges of BNNRs. Single-edge (with all its different possibilities, for example, for a BNNR with 2-line PHLD at single-edge there are 8 possibilities) as well as both-edge passivations have been considered for all the ribbons in this study by passivating each edge atom with hydrogen atom. Density of states (DOS) and projected-DOS (pDOS) analysis have been accomplished to understand the underlying reason for various properties. We find that passivation lead to different effects on the electronic and magnetic properties of a system, and the effects are mainly based on the line-defect introduced and/or on the atoms which are present at the passivated edge. In general, we find that, passivation can play a key role in tuning the properties of a system only when it has a zigzag edge.

## 1. Introduction

Among the analogues of graphene nanoribbons (GNRs) [1], boron nitride nanoribbons with zigzag edges have attracted huge attention in the recent years (zBNNRs) as they are shown to be potential candidates for electronic and spintronic devices [2-4]. Several studies have shown that the exciting electronic and spintronic properties of zBNNRs are due to their edge states [2-6], and mainly, when these edge states have been tuned by external factors like application of electric-field [4], doping [7], passivation [2, 3, 5, 8] etc. Among these external factors, effects of passivation on electronic and magnetic properties of BNNRs have been studied extensively using several theoretical methods. Such studies have shown that BNNRs can be tuned from insulating to semi-conducting to metallic either for one spin channel or for both spin-channels, depending on the nature/amount of passivation. But, majority of these studies have concentrated on BNNRs with zigzag edges which are hard to achieve experimentally, even for GNRs [9, 10].

General experimental procedures like etching [11], chemical vapor deposition [9, 10] etc. will generally produce defects in the otherwise perfect ribbons. Among the several kinds of defects, point defects and stone-wales defects are found to be ubiquitous in both graphene and boron-nitride sheets [9-11]. Though there are calculations on the effects of stone-wales defects on the electronic and magnetic properties of zBNNRs [5], there are very few studies [12] on the effects of stone-wales line-defects (also generally called as pentagon-heptagon line-defects). Recently, our group has studied the effects of odd and even-line PHLDs on the electronic and magnetic properties of BNNRs [6], and very recently, Tang et.al [13] have considered hybrid BNNRs (joined by an odd-line PHLD) also for such studies. Our study [6] already revealed that pristine zBNNRs with PHLDs have potential applications in electronic devices. In this study, we would like to understand the effects of hydrogen passivation on these zBNNRs with PHLDs.

Previous studies on 10-zBNNR with PHLDs [6] have shown that (i) the electronic and magnetic properties of the ribbon majorly depend on the nature of the PHLD (i.e. odd-line PHLD or even-line PHLD) rather than its size (i.e. 2-line or 6-line) and (ii) the amount of the energy required to add a PHLD in a 10-zBNNR increases with the size of the defect. Thus, we have only considered the 1-line-PHLD (1LD, figure 1b) and 2-line-PHLD (2LD, figure 1c) as representative candidates for odd and even-line PHLDs, respectively. In this work we have investigated the structural, electronic and magnetic properties of passivated zigzag boron-nitride nanoribbons (zBNNRs) with pentagon-heptagon (PH) line-defects (LDs). We have shown that, it is possible to tune in the properties of passivated zBNNRs, from insulating – semiconducting – metallic, depending on the position, passivation and number of lines of the PH-line-defect present in the zBNNR ribbon. Rest of the article is arranged as follows: First, we will discuss the computational details. Next, the results are presented and discussed followed by some important conclusions.

## 2. Computational Details

Spin-polarized Density Functional Theory (DFT) calculations have been performed for all the systems, using the ab-initio software package SIESTA [14, 15]. The generalized-gradient-approximation (GGA) with the Perdew-Burke-Ernzerhof (PBE) [16, 17] is used for the exchange-correlation functional and double-$\zeta$ basis sets are used to expand wave-functions. Interaction between the ionic cores and the valence electrons is accounted by the norm conserving pseudo-potentials [18] in the fully non-local Kleinman-Bylander form [19]. To represent the charge density, a reasonable mesh-cut-off of 300 Ry is used for the grid integration. Systems are considered to be optimized (conjugate-gradient method is used) till the forces acting on all the atoms are less than 0.04 eV / Å. Brillouin-zone is sampled by 36 × 1 × 1 k-points, using the Monkhorst-Pack scheme[20], for the full-relaxation of the geometry, and 96 × 1 × 1 k-points, for the calculations of electronic and magnetic properties. Cubic unit-cells with the initial lattice vectors (4.932, 0, 0); (0, 35, 0) and (0, 0, 15) Å, have been considered for all the systems of width ~ 2 nm. After optimization, all the systems remained flat and a change in the lattice-vectors has been found, mainly, along the X-direction from 4.932 to ~ 5.01 Å. A vacuum of 15 Å has been considered in the non-periodic directions in all the calculations to avoid any spurious interactions between the nanoribbons and their periodic images. A broadening parameter of 0.05 eV has been used while plotting DOS and pDOS plots.

In all our calculations, PHLDs have been introduced either at a single edge or at both the edges of 10-zBNNRs (here after, NRs). Passivation has been performed only at the edges, in single edge and in both edges. Systems have been named considering the passivation and position of the LD as shown in figure 1. Names and structures of all the systems are given in figures S1-S3, for clarity. Spin-polarized first-principles calculations have been performed on all these systems with different spin-configurations. Results of (UU, DD) spin-configuration are presented first and are then compared with other spin-configurations. Here U/D represents up/down-spin and the first/second element in the order pair, (UU, DD), represents the spin at Boron/Nitrogen edge.

## 3. Results and Discussions

In this section, first we will present the results of formation-energy ($E_{Form}$) and spin-polarization ($S_{pol}$) of all the systems and then we will discuss their electronic and magnetic properties. While presenting the results, for an easy understanding, 10-zBNNRs with 1 and 2-line PHLDs have been categorized as (a) Systems without defect (b) Systems with 1-line PHLD at one edge (c) Systems with 1-line PHLD at both edges (d) Systems with 2-line PHLD at one edge and (e) Systems with 2-line PHLD at both edges. Values of formation energy and spin-polarization are given in Table 1 in this manner. Here, we have presented the complete results of the passivated systems and also we've compared them with the results of pristine systems. [6]

*3.1. Formation energy and Spin-polarization*

Formation-energy ($E_{Form}$) of a system is calculated as:

$$E_{Form} = [E_{tot} - nB * E_B - nN * E_N - nH * E_H] / [nB + nN + nH] \qquad (1)$$

where, $E_{tot}$ is the total energy of the system; $E_B$, $E_N$ and $E_H$ are the total-energies per atom of α-boron, $N_2$-molecule and $H_2$-molecule, respectively. nB, nN and nH are the number of boron, nitrogen and hydrogen atoms in the system, respectively. $E_{Form}$ values of all the systems are negative suggesting that all these systems are feasible thermodynamically. Some of the important trends which we have observed from the table 1 are: (i) The stability order is always "both-edge passivated systems > N-edge passivated systems > B-edge passivated systems > pristine systems" for all the systems studied, and this order clearly shows that passivation increases the stability of a system irrespective of whether it has perfect edges or defect edges or both. This is expected because passivation removes dangling bonds in the systems, and hence, they acquire stability. (ii) Irrespective of whether the edge is perfect or defect, passivating a zigzag-edge with nitrogen atoms is more stable than passivating a zigzag-edge edge with boron atoms. This is because passivating the edge boron atoms gives a sextet-configuration to each boron atom, whereas, for nitrogen it will provide an octet-configuration, and hence, N-ed-pa is more stabilized (please see SI for further information). (iii)  Among the passivated ribbons, perfect-edge passivated ones are always more stable than the defect-edge passivated ones. This is because of two facts (a) passivating a zigzag edge gives more stability than passivating an armchair edge and (b) there will be loss in energy with the introduction of a defect. 1LD-NRs will have armchair-edges, and hence, passivation can't bring huge stability. 2LD-NRs will have zigzag-edges similar to perfect NRs, but, the gain in energy through passivation is not enough to compensate the loss in energy due to the introduction of 2LD.

Next, we have calculated the spin-polarization of these systems using the formula, "$S_{pol} = Q_{up} - Q_{down}$", where, $Q_{up}$ is the Spin-up charge density and $Q_{down}$ is the spin-down charge density.  Values of the $S_{pol}$ have been given (only for (UU, DD) spin-configuration) in the table 1. In general, it is known that non-passivated zigzag edges can lead to finite spin-polarization [6]. Thus, when the spins at the two different edges are not interacting (as in our case), we can expect a finite spin-polarization only when the spins at both the edges are of same kind ( like (UU, UU) etc.) or when the spin-moment at one edge is completely destroyed while (either through "passivation" or through the "introduction of odd-line defect" or by having "UD or DU type spin-configuration") keeping the other edge's spin-moment finite. $S_{pol}$ values in table 1 (for any system) can be understood based on the above notes. For clarity, reasons for all the $S_{pol}$ values and stability orders of table 1 are given in SI.

*3.2. Electronic and magnetic properties:*

In this section, we have used DOS and pDOS plots to understand and explain the electronic and magnetic properties of all the systems.

3.2.1. Systems without defect: pDOS plots of these systems are shown in figures 2a-2d. In agreement with the previous works [6], we find that bare 10-zBNNR (B-NR-N) are *anti-ferromagnetic half-metals*. As this half-metallic nature is due to the edge-states [6], systems have transferred to semi-conductors (when only one edge is passsivated, figures 2b, 2c) or insulators (when both edges are passivated, figure 2d) depending on the number of edges being passivated. Interestingly, in the pDOS of bare-ribbon (figure 2a) we find that the gap between the nitrogen edge-states, across the Fermi level, is less compared to the boron edge-states. In chemistry, it is well known that [21, 22] higher the HOMO-LUMO gap of a compound higher is its chemical stability. By analogy, we should expect that N-edge states should be more reactive than B-edge states and this is what is reflected in the higher stability of "B-NR-NH" system than "HB-NR-N" system (see table 1 and section 3.1). A further inspection on the edge states is given in SI.

3.2.2 Systems with 1-line-PH-defect at single edge: From the pDOS plots (figures 2e to 2l) it is apparent that, all these systems are either insulating or semi-conducting. From the previous studies, [2, 4, 6, 23] already we know that, BNNRs with armchair edges are always non-magnetic and are insulating irrespective of their passivation. Also, these studies reveal that, BNNRs with at least one bare zigzag edge only can be magnetic. Here also, after reconstruction, one of the two edges has changed its edge nature from zigzag to armchair, and hence, the spin-polarized states of this edge have been lost (figures 2e, 2f). From the pDOS plots, we find that the semi-conducting nature of these systems is because of the free-electron states at the bare edges, and, mainly because of the bare zigzag-edge. Once this bare zigzag-edge is passivated, these edge states are immediately vanished, and lead to a change in the system's property from semi-conducting to insulating. It is nice to find that, unlike the perfect edge, defect edge passivation has very less impact on the pDOS across the Fermi level (compare figures 2e, 2k and figures 2f, 2l). Again, we find that N-edge passivation leads to more band-gap than B-edge passivation in agreement with the results of perfect-ribbons. Importantly, we find that B-edge reconstruction has less impact on the band-structure compared to N-edge reconstruction and, similar to passivation. Thus, not only passivation but also reconstruction can change the system's behavior, and it is the zigzag edge of the system which dictates the system's electronic and magnetic.

3.2.3. Systems with 1-line-PH-defect at both edges: pDOS plots (see figure 3) shows that all the systems are insulating irrespective of the edge passivation and, as discussed in the sub-section 3.2.2, this is because of the armchair nature of the edges in these systems. Thus, we understood that, reconstruction and passivation suppress the magnetic behavior of the system.

3.2.4. Systems with 2-line-PH-defect at one edge: pDOS plots of these systems show that (see figures 4a to 4h) they can be tuned from half-metallic to semiconducting to insulating by changing the passivation at a specific edge. When both the edges are bare, they possess finite $S_{pol}$ across the Fermi level similar to the "systems without defect". But, unlike the "systems without defect", these systems have two types of zigzag edges, namely, perfect zigzag edge and defect zigzag edge, and, as can be seen from the figures 4a to 4h, the properties of the system differ based on whether we passivate a perfect zigzag edge or a defect zigzag edge. Thus, in these systems, we have more freedom in tuning the properties than in the "systems without defect". Except for "HB-NR-2LD-N" system, all the systems have transformed from half-metallic to semiconducting, when one of the edges is passivated. As B-edge passivation has less impact on the bandstructure than the N-edge passivation, half-metallic behavior of "HB-NR-2LD-N" is expected. This further shows that, major changes in the bandstructure can be found only when the N-edge states are passivated, irrespective of whether the N-atoms are at defect edge or at the perfect edge. Finally, both-edge passivated systems are insulating as there are no more free-electron states.

3.2.5. Systems with 2-line-PH-defect at both edges: As both the edges are zigzag, these systems also behave as *anti-ferromagnetic half-metallic* systems, when their edges are bare (see figure 4i). Again, single-edge passivation change their electronic behavior from half-metallic to semiconducting and passivation of boron-edge didn't bring any great difference in the pDOS compared to the passivation of nitrogen edge (see figure 4j and 4k). Also, pDOS of the system in figure 4k compares well with that of the system in figure 4e, as both of them have N-atoms at the bare-defect-edge, although both of them have different B-edges (4k has defect B-edge and 4e has perfect B-edge). This again proves that B-edge reconstruction has less effect on the bandstructure than the N-edge reconstruction. Thus, compared to 1-line-PH-defect systems, 2-line-PH-defect systems have more ability to tune their electronic properties which in turn arise because 2-line-PH-defect keeps the zigzag edge nature.

3.2.6. Effect of Spin-configuration: As the bandstructure and $S_{pol}$ are mainly be affected by the nature of the bare zigzag-edge (see previous sections), to understand the effects of spin-configuration, here, we have considered the systems with boron-atoms at bare-zigzag-edge. Among such systems, some selected ones have been shown in figure 5 (others show exactly the same trend) with (UU, DD), (UU, UU) and (UD, UU) spin-configurations. Clearly, the only change from the (UU, DD) to (UU, UU) configuration is that those edge states (either of boron or nitrogen) which are as the up-spin have changed to the down-spin, just like a reflection across the Fermi level. This is because, there is no interaction between the spins across the edges. But, there are differences between the configurations (UU, DD) [or (UU, UU)] and (UD, UU). Also, these differences dependent on whether the boron atoms are at the perfect zigzag-edge or at the defect-zigzag-edge. In perfect-edge systems (figures 5a and 5b), the changes in the B-edge states between (UU, UU) and (UD, UU) configurations is due to the spin-symmetry which we are bringing (in the latter system) and this is the reason for the spin-unpolarized DOS of B-edge states near the Fermi level in (UD, UU) system. In defect-edge systems (figure 5c), though there is spin-symmetry, as the edge has 5 and 7 membered ring atoms (which are different by their nature), there is spin-polarization near the Fermi level. Please see section 3.2.6 of SI and ref [6] for further understanding. Based on the above results, we can easily expect the changes in the bandstructure of a system (width > 2 nm) with a change in spin-configuration, if we know (i) system's bandstructure for a particular spin-configuration and (ii) the edge information (i.e. defect or perfect).

## 4. Conclusions

In conclusion, all the properties of the systems which have been presented here are mainly dependent on the edge-nature of the ribbon and this edge nature can be tuned using PH-line-defect number and passivation as tools. We have shown that, for (UU, DD) spin-configuration, a system with, (i) both the edges of armchair nature will behave as non-magnetic insulator, irrespective of passivation, (ii) one zigzag edge and one armchair edge will behave as spin-polarized semi-conductor either when both the edges are bare or when the defect-edge is passivated, and as non-magnetic insulator either when both the edges are passivated or when the perfect-edge is passivated and (iii) both the edges of zigzag nature will behave as anti-ferromagnetic half-metals when both the edges are bare; spin-polarized semi-conductors when a single-edge is passivated; and non-magnetic insulator when both the edges are passivated. We have also shown that, for all the systems, the B-edge passivation and reconstruction has less effect on the bandstructure than the N-edge and the root cause for the change in the properties of the system with a change in spin-configuration at the edges has been understood as the difference in the edge atom's properties for the defect and perfect edge. With passivation, we obtained the stability order as: both-edge passivated > single-edge passivated > pristine, for all the systems and we have shown that N-edge passivation is more stable than B-edge passivation (irrespective of whether the edge is defect or perfect) and perfect-edge passivation is more stable than defect-edge passivation, with reasons.

**Acknowledgements**

We thank TUE-CMS, JNCASR for providing the computational facilities. SKP thanks DST, India for funding.

**Table 1:** Formation energy ($E_{form}$) in eV/ atom and spin-polarization ($S_{pol}$) for all the systems in (UU, DD) configurations are given. System names are given according to the nomenclature explained in the section 3.

| S.No. | System name | $E_{Form}$ (eV/atom) | $S_{pol}$ |
|---|---|---|---|
| *Systems without defect* | | | |
| 1 | B-NR-N | -1.352 | 0.000 |
| 2 | B-NR-NH | -1.547 | -2.000 |
| 3 | HB-NR-N | -1.514 | -2.000 |
| 4 | HB-NR-NH | -1.694 | 0.000 |
| | *Systems with 1-line-PH-defect at Single edge* | | |
| 5 | BN-1LD-NR-B | -1.355 | 2.000 |
| 6 | BN-1LD-NR-N | -1.359 | 1.983 |
| 7 | BN-1LD-NR-BH | -1.518 | 0.000 |
| 8 | BN-1LD-NR-NH | -1.555 | 0.000 |
| 9 | HBN-1LD-NR-B | -1.469 | 2.000 |
| 10 | HBN-1LD-NR-N | -1.465 | 2.000 |
| 11 | HBN-1LD-NR-BH | -1.617 | 0.000 |
| 12 | HBN-1LD-NR-NH | -1.647 | 0.000 |
| | *Systems with 1-line-PH-defect at both edges* | | |
| 13 | BN-1LD-NR-1LD-BN | -1.365 | 0.000 |
| 14 | HBN-1LD-NR-1LD-BN | -1.477 | 0.000 |
| 15 | HBN-1LD-NR-1LD-BNH | -1.574 | 0.000 |
| | *Systems with 2-line-PH-defect at Single edge* | | |
| 16 | N-2LD-NR-B | -1.254 | 0.000 |
| 17 | B-2LD-NR-N | -1.241 | 0.000 |
| 18 | N-2LD-NR-BH | -1.421 | -2.000 |
| 19 | B-2LD-NR-NH | -1.440 | -2.000 |
| 20 | HN-2LD-NR-B | -1.445 | 2.000 |
| 21 | HB-2LD-NR-N | -1.407 | 2.000 |
| 22 | HN-2LD-NR-BH | -1.596 | 0.000 |
| 23 | HB-2LD-NR-NH | -1.591 | 0.000 |
| | *Systems with 2-line-PH-defect at both edges* | | |
| 24 | B-2LD-NR-2LD-N | -1.144 | 0.000 |
| 25 | B-2LD-NR-2LD-NH | -1.339 | 2.000 |
| 26 | HB-2LD-NR-2LD-N | -1.314 | -2.000 |
| 27 | HB-2LD-NR-2LD-NH | -1.494 | 0.000 |

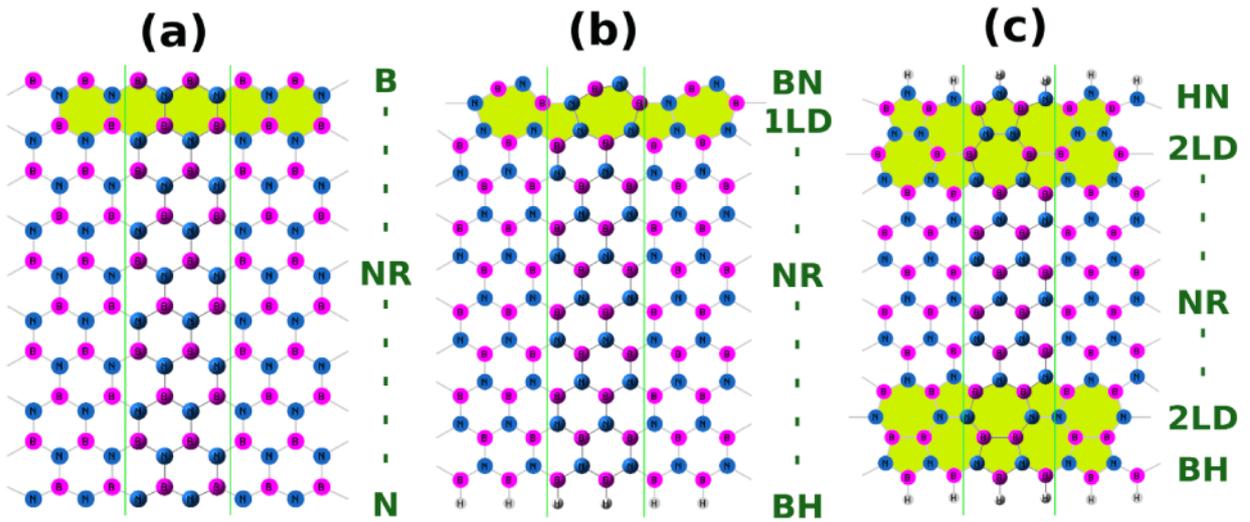

**Figure 1:** Example systems illustrating the nomenclature used in the present study. (a) pristine 10-zBNNR (B-NR-N) (b) perfect edge (with boron atoms) passivated 10-zBNNR with 1-line PHLD at one of its edges (BN-1LD-NR-BH) and (c) both edge passivated 10-zBNNR with 2-line PHLDs at both of its edges.

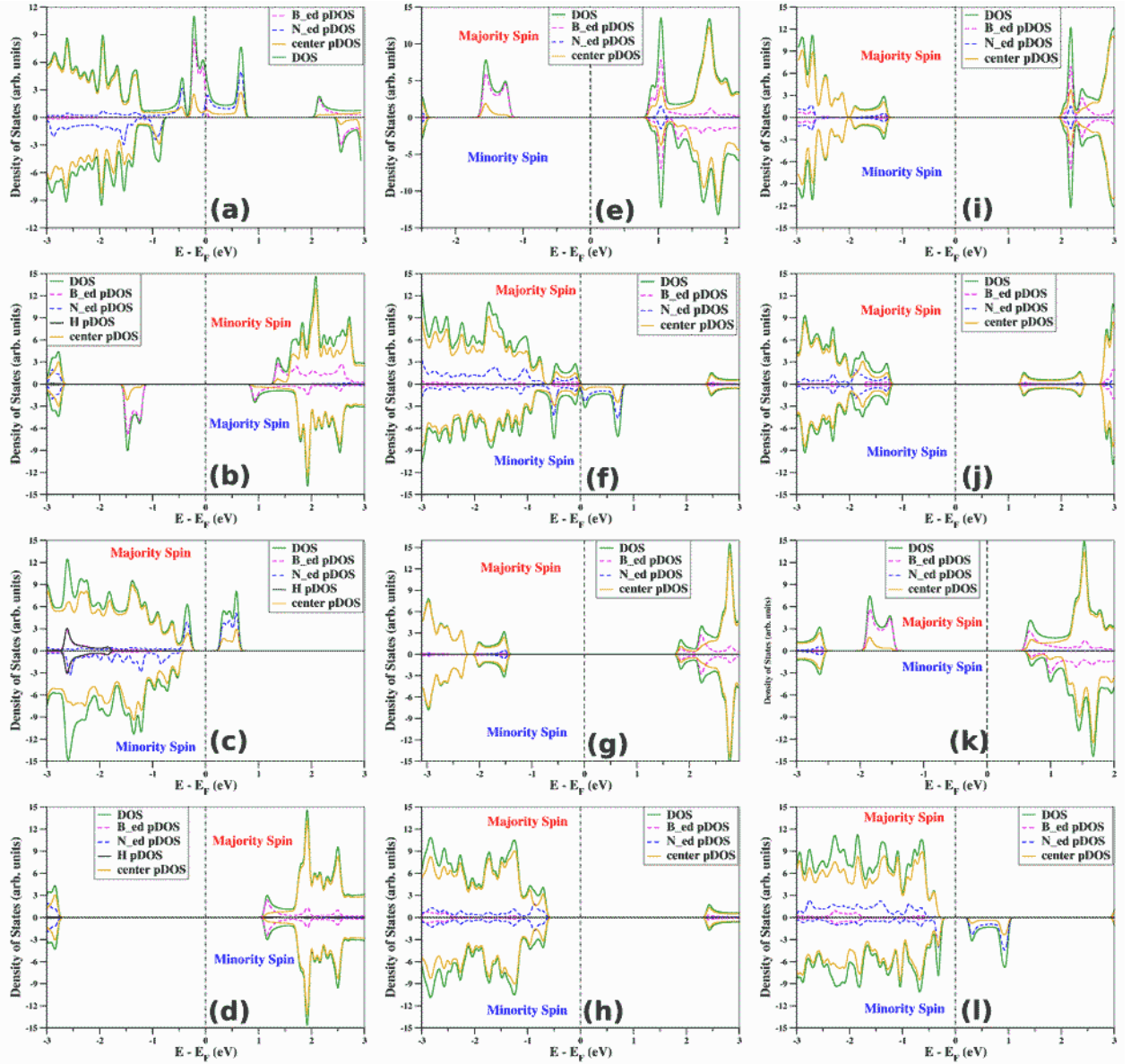

**Figure 2:** pDOS plots of (a) B-NR-N (b) B-NR-NH (c) HB-NR-N (d) HB-NR-NH (e) BN-1LD-NR-B (f) BN-1LD-NR-N (g) HBN-1LD-NR-BH (h) HBN-1LD-NR-NH (i) BN-1LD-NR-BH (j) BN-1LD-NR-NH (k) HBN-1LD-NR-B (l) HBN-1LD-NR-N. Up-spin pDOS is shown at the top and down-spin's at the bottom of these plots. Majority and Minority-spins are defined based on $S_{pol}$ (see SI).

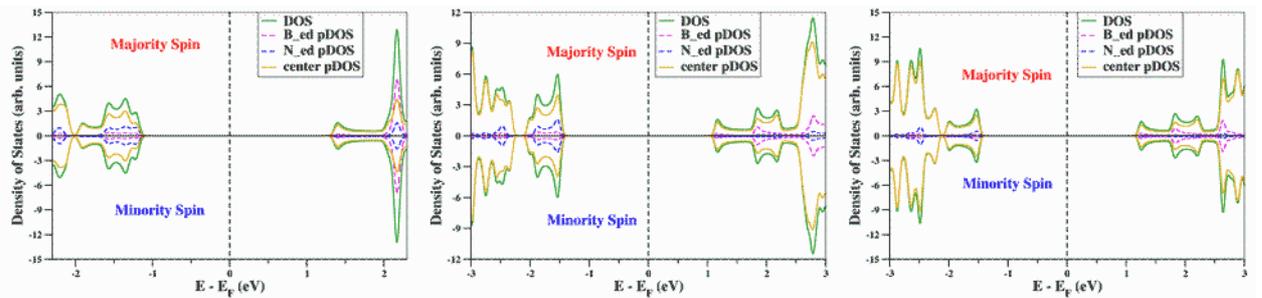

**Figure 3:** pDOS plots of (a) BN-1LD-NR-1LD-BN (b) BN-1LD-NR-1LD-BNH (c) HBN-1LD-NR-1LD-BNH

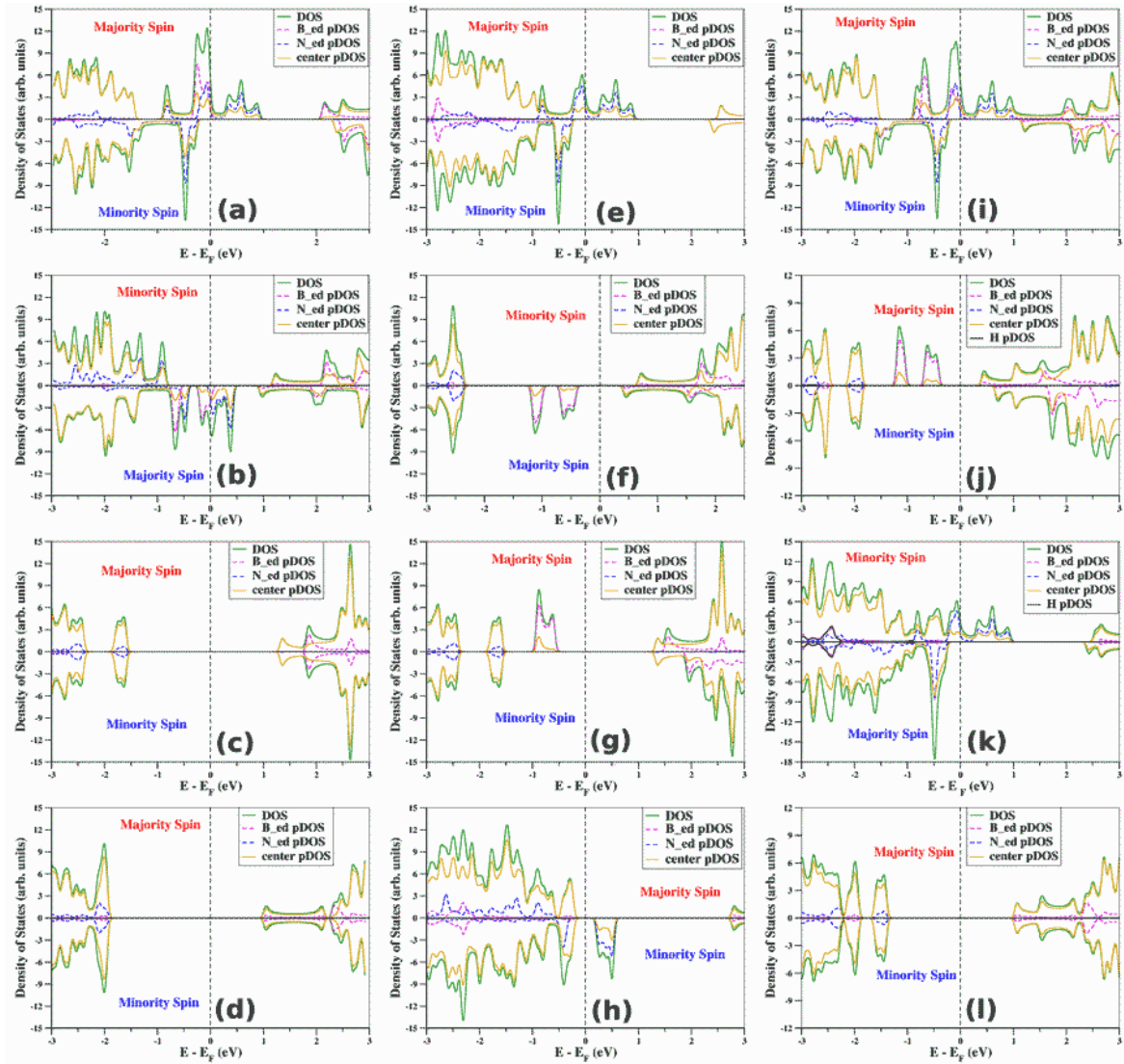

**Figure 4:** pDOS plots of (a) N-2LD-NR-B (b) B-2LD-NR-N (c) HN-2LD-NR-BH (d) HB-2LD-NR-NH (e) N-2LD-NR-BH (f) B-2LD-NR-NH (g) HN-2LD-NR-B (h) HB-2LD-NR-N (i) N-2LD-NR-2LD-B (j) HN-2LD-NR-2LD-B (k) N-2LD-NR-2LD-BH (l) HN-2LD-NR-2LD-BH

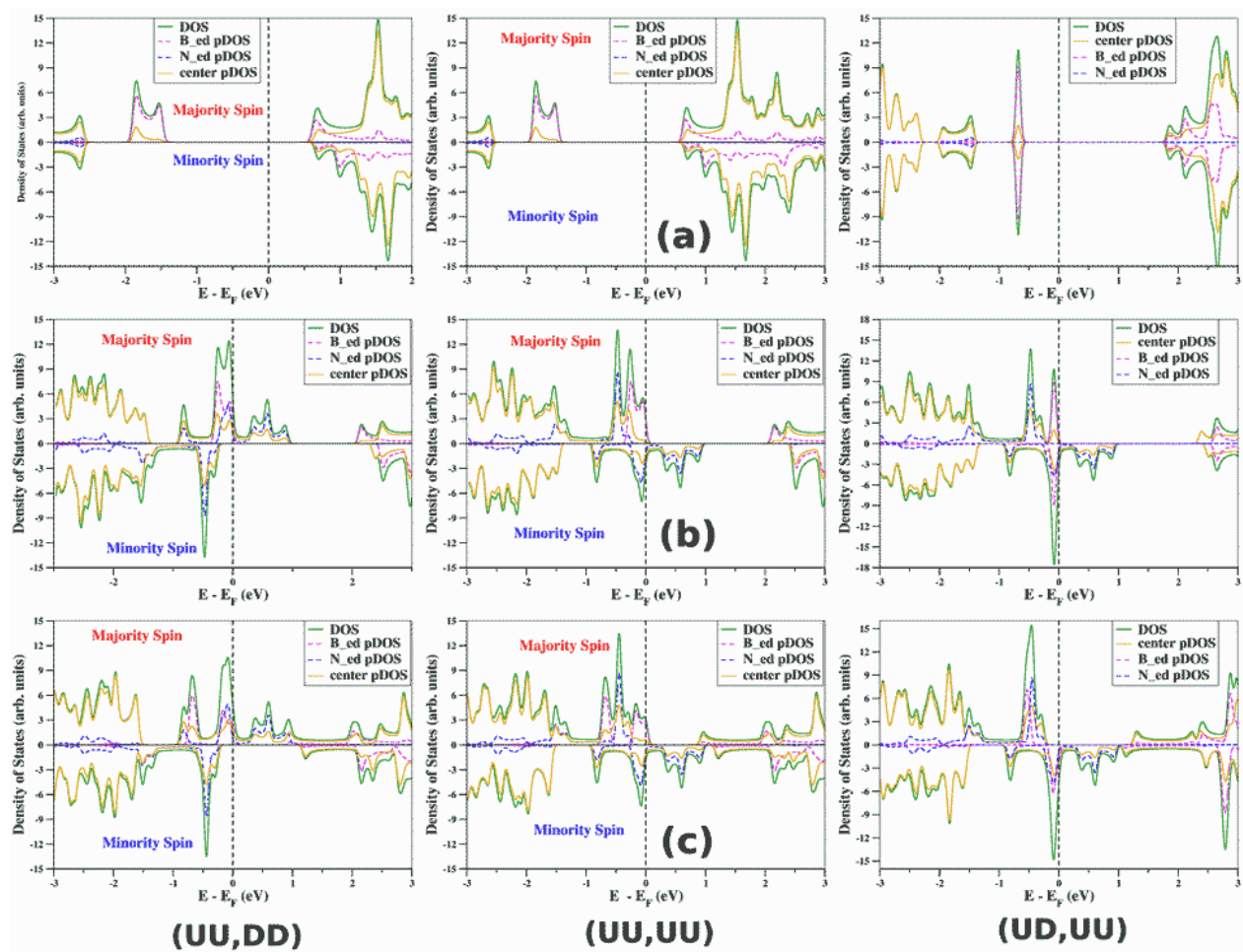

**Figure 5**: pDOS plots of (a) HBN-1LD-NR-B (b) N-2LD-NR-B and (c) N-2LD-NR-2LD-Bin all the three different spin-configurations. Please, observe how the B-edge states are changing with spin-configuration in all the three cases.

# Supporting information for

# "Effects of Edge Passivations on the Electronic and Magnetic Properties of Zigzag Boron-Nitride Nanoribbons with Even and Odd-Line Stone-Wales (5–7 Pair) Defects"

## SRKC Sharma Yamijala and Swapan K Pati

**Spin-configurations considered in our study:**

For all the systems, we have performed the spin-polarized DFT calculations. Based on the previous studies, [1-4] we considered 4-spin configurations (UU, UU), (UD, UU), (UU,DD) and (DD, UU), where U and D represents up and down-spins, respectively; and the first and second element in an ordered pair represents the B and N-edges, respectively (We found the spin-moment as zero for arm-chair edges, and hence, not mentioned here). As (UU, DD) and (DD, UU) gave same results (except for the change that spin-up in one system is spin-down in the other and vice-versa) for all the properties calculated in this work, we referred (UU, DD) for comparisons. Also, we find that the configurations (UU, UU) and (UU, DD) have similar energies ( < 10 meV, which is near the error bar of our calculations). Finally, in agreement with the previous DFT studies [1-5] we found that (UD, UU) configuration as the stable configuration among all the systems and (UU, DD) has shown interesting results. As the number of systems is large and as (UU, DD) gave interesting results we've first presented the results with (UU, DD) configuration. For completeness, results of other spin-configurations have also been compared (see main-article). Indeed, we find that it is possible to guess the results of the other spin-configurations based on the results explained for the (UU, DD) configurations for 10-zBNNR systems as explained in our previous work. [5]

**Structures of all the systems considered in this study:**

Naming of the systems is done as shown in Figure 1 of the main article. For example, in system HB-2LD-NR-NH, "2LD-NR" denotes 10-zBNNR has a 2-line PHLD at one of its edges and 'HB' before 2LD denotes that the defect edge has boron atoms and are passivated with hydrogen atoms. Finally, 'NH' after NR denotes the perfect edge has nitrogen atoms and is passivated. Similarly, we have used the indices "N", "B" and "BN" to convey that the edges are not passivated and they have nitrogen, boron and "both boron and nitrogen (i. e. its an armchair edge)" atoms, respectively.

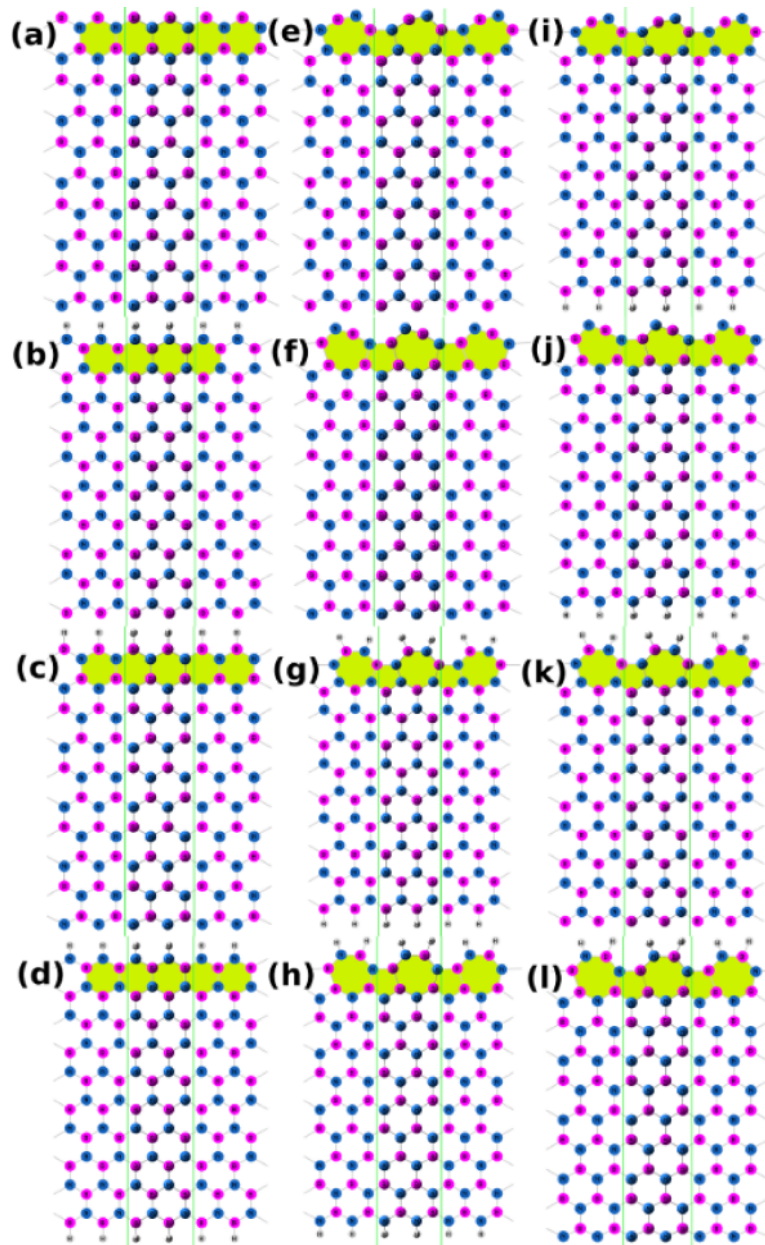

**Figure S1**: Structures of (a) B-NR-N (b) B-NR-NH (c) HB-NR-N (d) HB-NR-NH (e) BN-1LD-NR-B (f) BN-1LD- NR-N (g) HBN-1LD-NR-BH (h) HBN-1LD-NR-NH (i) BN-1LD-NR-BH (j) BN-1LD-NR-NH (k) HBN-1LD-NR- B (l) HBN-1LD-NR-N

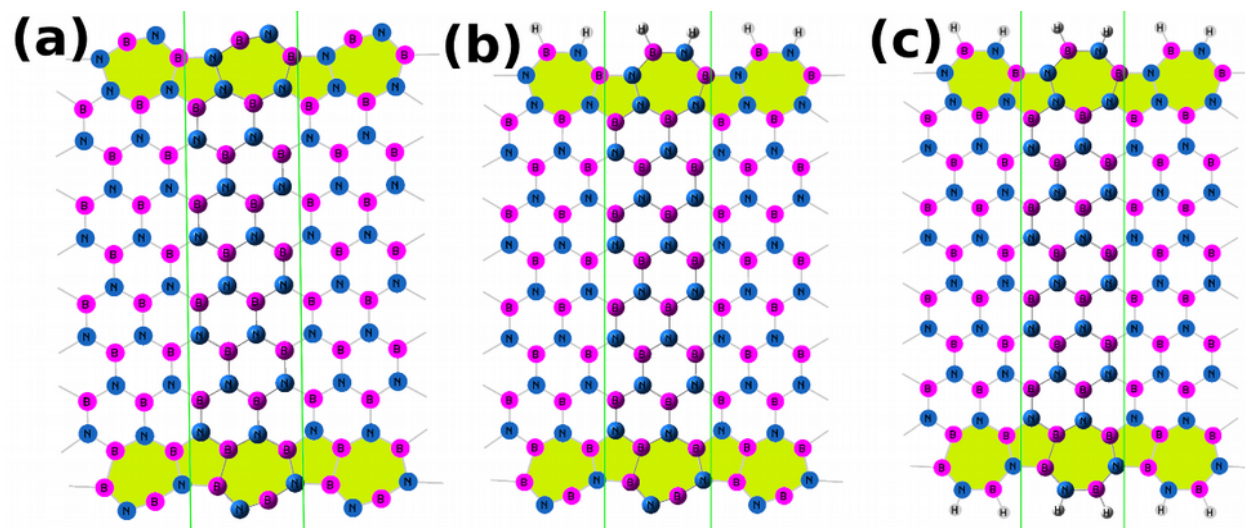

**Figure S2**: Structures of (a) BN-1LD-NR-1LD-BN (b) BN-1LD-NR-1LD-BNH (c) HBN-1LD-NR-1LD-BNH

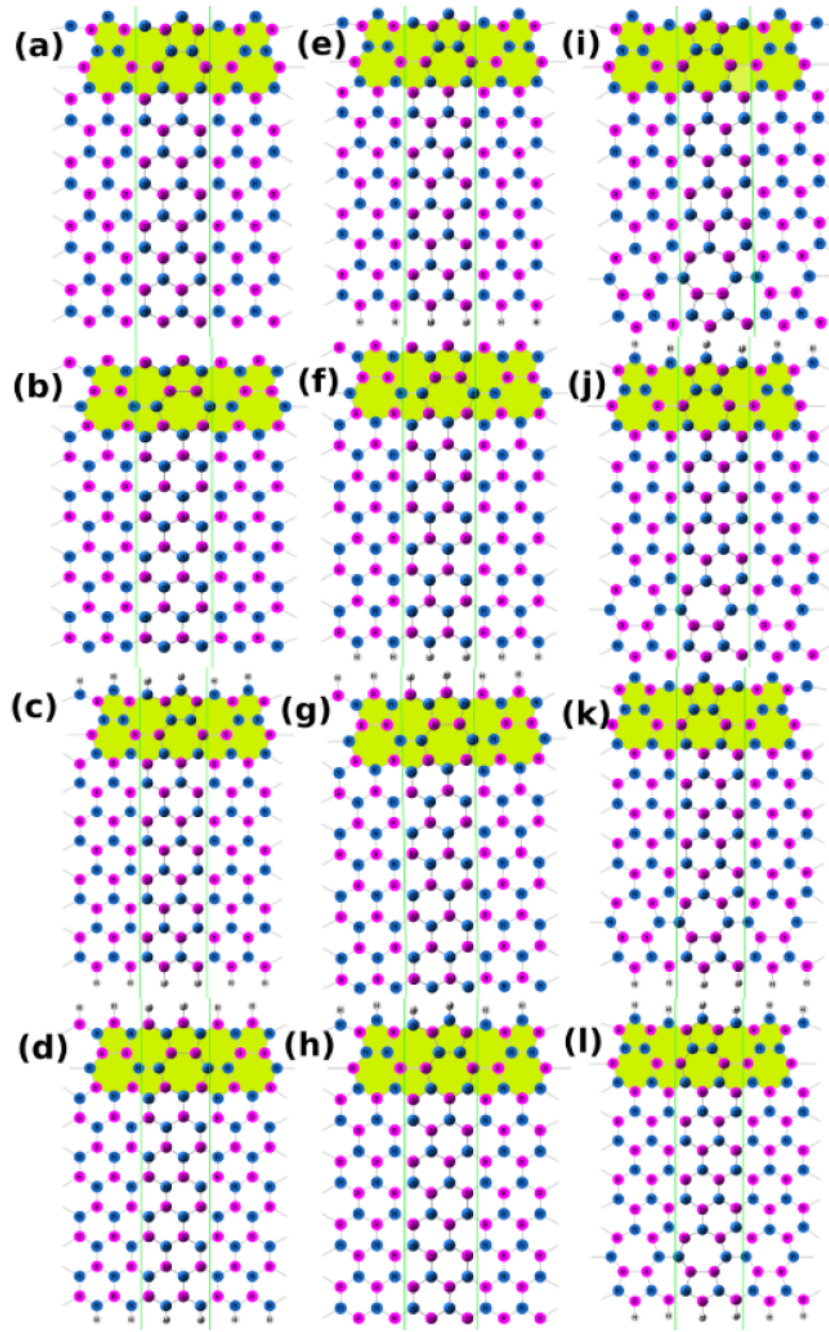

**Figure S3**: Structures of (a) N-2LD-NR-B (b) B-2LD-NR-N (c) HN-2LD-NR-BH (d) HB-2LD-NR-NH (e) N-2LD- NR-BH (f) B-2LD-NR-NH (g) HN-2LD-NR-B (h) HB-2LD-NR-N (i) N-2LD-NR-2LD-B (j) HN-2LD-NR-2LD-B (k) N-2LD-NR-2LD-BH (l) HN-2LD-NR-2LD-BH

### *3.1. Formation energy and spin-polarization*

### 3.1.1. Systems without defect

Under this section, always, we will have four systems viz., pristine, B-edge passivated, N-edge passivated and both-edge passivated ribbons. All of them are shown in the first column of figure S1 and, as mentioned earlier, the results are discussed only for the (UU, DD) configuration. From the $E_{Form}$ values (see rows 1 to 4 in table 1 of main article), it is clear that the stability of the systems are in the order of, both-edges H-passivated (Bo-ed-pa) > N-edge passivated (N-ed-pa) > B-edge passivated (B-ed-pa) > pristine (pr) ribbons.

Octet vs Sextet configuration:

Firstly, it is interesting to notice that, there is a difference in the stability of the systems based on whether the passivated edge has nitrogen or boron atoms and the difference between them is greater than room-temperature (see table 1). The reason for this stability can be explained as follows. As the edge atoms of BNNRs are $sp^2$ hybridized, there will be three $sp^2$-hybrid orbitals with one electron each and a completely filled (vacant) $p_z$-orbital for the edge-nitrogen (boron) atom. Out of these three $sp^2$-hybrid orbitals two will be bonded to inner BNNR network and one will be half-filled and this orbital will be ready to form a bond (here with the hydrogen). For the case of boron, although the bond formation of this half-filled orbital with hydrogen helps to passivate its dangling bond, it can only give a sextet-configuration (6-electrons in the valence-shell, because the $p_z$-orbital is empty). Whereas, for the case of nitrogen, bond formation not only passivates the dangling bond but also provides an octet-configuration, and hence, N-ed-pa is more stabilized.

"Availability of lone-pair" is the other reason for higher stability of HN-NR-B systems:

Clearly, the explanation of octet formation is satisfactory only when there is one free-electron in the atom used for passivation. For the case of hydrogen, this condition will be satisfied in its atomic or radical forms. But, if we consider the hydrogen in its molecular form (i.e. $H_2$), then the question of how easily a $H_2$-molecule will dissociate to form a bond with the edge-atoms of BNNR will arise. We took the help of molecular-orbital-theory [6] (MOT) to answer the above question, and we found that, even in $H_2$-environment N-ed-pa is more stable than B-ed-pa and the explanation is given below.

According to MOT, we know that $H_2$ molecule has one bonding orbital (BO) and one anti-bonding orbital (ABO) and its bond-order (i.e. [no.of electrons in BO – no.of electrons in ABO]/2) is one. When a $H_2$-molecule is brought near to the N-ed, the nitrogen atom at the edge will try to donate its lone-pair (not the electrons in the $sp^2$-hybrid orbitals because they will have more electro-negativity compared to $p_z$-orbitals, and hence, difficult to donate) to the ABO of the $H_2$-molecule. This donation (even if we consider it as very less) will surely increase the electron-population to the ABO, and hence, decreases the bond-order of the $H_2$-molecule. Decrease in the bond-order means, now $H_2$ can be easily dissociated than before, and thus, can form a new bond with nitrogen. Boron, due to absence of this lone-pair in its $p_z$-orbital might not dissociate $H_2$ easily and this could be the reason for the poor stability of B-ed-pa in $H_2$-environment. Thus, in a BNNR, both edges passivation will make the system more stabilized as it will remove all the dangling bonds in the system. But, if we want to leave only one edge as bare and also if we want the system to be more stable, then *it is the B-ed which we have to keep bare*, for the reason explained above. Also, please see the discussion on the N-edge states and their reactivity in the section 3.3.1.

### 3.1.2. Systems with 1-line-PH-defect at single edge

Under this section, we will always have eight systems and all the systems are shown in the second and last columns of figure S1. $E_{Form}$ values (see table 1) clearly show the stability order of these ribbons as: HBN-1LD-NR-NH > HBN-1LD-NR-BH > BN-1LD-NR-NH > BN-1LD-NR-BH > HBN-1LD-NR- B > HBN-1LD-NR-NBN-1LD- NR-N ≈ BN-1LD-NR-B. In agreement with our understanding on without-defect-ribbons, here also, we observe that, the edge passivated systems are more stabilized than the pristine ribbons; Both-edge-ribbons (Bo-ed-pa) ribbons are more stabilized than single-edge-passivated (si-ed-pa) ribbons and N-ed-pa ribbons are more stable than B-ed-pa ribbons (reasons are same as that of without-defect-ribbons).

Apart from the above observations, one can also see that "perfect-edge passivated ribbons are more stable than defect-edge passivated ribbons". We can explain this based on the following facts: 1) Not only the passivation, but also the 5-7-edge-reconstruction (also called as self-passivation in the literature, and it changes the zigzag edge to armchair), will gives stability to the bare zGNRs [7]. 2) Bare arm-chair edges are more stable than the bare zigzag-edges, in BNNRs [8]. Thus, by introducing a 1-line-PH-defect, we force the system to change its edge nature from zigzag to armchair, and also, unknowingly, gave stability to this defect edge. Now, as the edge with 1-line-PH-defect is already stabilized (because of its armchair nature), the gain in stability due to passivation is very less for this edge. On the other hand, *the gain in stability is more when the passivated edge is perfect*, because, perfect edge didn't lose its zigzag nature.

The above explained zigzag-edge nature is the cause for two other important observations! They are: the difference in the $E_{Form}$ values between the N-ed and B-ed systems is, a) negligible, when the ribbons are either pristine or when they are passivated only at the defective edge; b) greater than room temperature, either when only the perfect edge is passivated or when both-edges are passivated. The later (former) observation is because the zigzag edge is passivated (not passivated) in those systems.

By joining the results from the sub-sections 3.1.1 and 3.1.2 (of main article and SI), we see that the following conditions are important to attain *stable* single-edge passivated (also called as half-bare in the literature) BNNRs. They are: 1) passivate the N-ed rather than the B-ed, if the system's both edges are zigzag; 2) passivate the zigzag-edge rather than the armchair-edge, if the system's one edge is zigzag and the other one is armchair (here armchair-edge is the edge with 1-line-PH-defect) and 3) *passivate the perfect edge and keep it with nitrogen atoms*, when the system has both zigzag and armchair edges (conclusion 3 is drawn from the conclusions 1 and 2).

### 3.1.3. Systems with 1-line-PH-defect at both edges

Under this section, we will always have three systems and all are shown in the middle column of figure S2. From table 1, the order of $E_{Form}$ values for these systems is: BN-1LD-NR-1LD-BN < BN-1LD-NR-1LD-BNH < HBN-1LD-NR-1LD-BNH. Again, in agreement with the above sub-sections, Bo-ed-pa ribbons are the most stable ones.

If we compare these systems with the systems in 3.1.1, we can see that (from $E_{Form}$ values), ribbons with defect at the edges are more stable than the systems with perfect edges (again because of edge-reconstruction) when both the edges are bare, which is in agreement with the previous studies [8, 9]. But, as the difference between them is ~ 0.01 eV, separating these systems from one another at room temperature might be difficult. On the other hand, if we compare either single or both edge passivated ribbons of this and 3.1.1 sections, then we can realize that, perfect edge systems are more stable (as expected, because of their zigzag-edge nature) than the defect edge systems and the energy difference between them is greater than 0.025 eV (Thus, based on $E_{Form}$ values, these systems should be able to exist separately at room-temperature).

### 3.1.4. Systems with 2-line-PH-defect at one edge

From this section on wards, we'll present the results of the ribbons with 2-line-PH-defect. In all these systems, both the edges are of zigzag type, but, still one can clearly distinguish a defective-zigzag-edge from a perfect-zigzag-edge. To our knowledge, nobody has studied these systems previously, and under this section we will always have eight systems and are shown in the first and second columns of figure S3. $E_{Form}$ values (see table 1) for these systems are in the order of: HN-2LD-NR-BH ≈ HB-2LD-NR-NH > HN-2LD-NR-B ≈ B-2LD-NR-NH > N-2LD- NR-BH > HB-2LD-NR-N > N-2LD-NR-B > B-2LD-NR-N. Clearly, we observed the expected order for $E_{From}$ values i.e. Bo-ed-pa > Si-ed-pa > pristine systems. The expected result in single-edge passivated systems is "B-2LD-NR-NH > N-2LD- NR-BH". The other result, " HN-2LD-NR-B > HB-2LD-NR-N", can also be expected if we remember that, HN-2LD-NR-B means that the system's perfect edge is with boron atoms and the defect edge is passivated (here, which is zigzag and have nitrogen atoms). But, these expected results in si-ed-pa systems lead to a new conclusion that, "*irrespective of whether the edge is perfect or defect, passivating a zigzag-edge with nitrogen atoms is more stable than passivating a zigzag-edge edge with boron atoms*".

One result which might need an explanation is: "Why N-2LD- NR-BH > HB-2LD-NR-N?" or, to put in another way, "Why the perfect-bare-edge is more stable than the defect-bare-edge, when both of them have nitrogen atoms at their bare-edge?" The reason is simple: "perfect-edge is always more stable than the defect-edge (2-line-PHLD), at least by 0.1 eV". This statement can be checked by comparing the $E_{From}$ values of systems in this and 3.1.1 sections. The less stability of the 2-line-PHLD edge compared to perfect edge could be because of the fact that 1) some extra energy is required to reconstruct the edge and 2) more importantly, this energy is not regained after the reconstruction. Later is because 2-line-PHLD didn't bring any change to the edge nature (here zigzag). In contrast, for the case of 1-line-PHLD, edge reconstruction changes the edge nature from zigzag to armchair (which is more stable), and hence, the energy required to reconstruct is compensated.

### 3.1.5. Systems with 2-line-PH-defect at both edges

Under this sub-section, we will always have four systems which are shown in the last column of figure S3. Observed order for the $E_{Form}$ (see table 1) is: N-2LD-NR-2LD-B < N-2LD-NR-2LD-BH < HN-2LD-NR-2LD-B < HN-2LD-NR-2LD-BHand the reason for this order can be understood based on our previous conclusions in sub-sections 3.1.1 – 3.1.4. Comparing the present systems with the systems in the section 3.1.4, we observe that systems with 2-line-

PHLD at one edge are more stable than 2-line-PHLD at both edges, and as the energy difference between them is more than room temperature, it should be possible to separate them at room temperature.

### 3.2. Spin-Polarization ($S_{pol}$)

Firstly, from the $S_{pol}$ values of table 1, we find that a 10-zBNNR with PHLDs have finite spin-polarization only when the below conditions are satisfied. (1) If the spin-configuration is of type (UU, DD) then it should have *exactly one zigzag edge* (irrespective of whether the zigzag edge is a perfect one or defect one) which is not passivated (2) If the spin-configuration is of type (UU, UU) then it should have *at least one zigzag edge* which is not passivated and (3) If the spin-configuration is of type (UD, UU) then it should have a *non-passivated zigzag edge* and the spins on this edge atoms should be interacting ferromagnetically. In the below, results are explained only for the (UU, DD) configuration and the reader is requested to use the same logic to understand the results of the other spin-configurations.

### 3. 2.1. Systems without defect

$S_{pol}$ values (see table 1) indicate that, whenever an edge is passivated, the spin moment which the edge was possessing previously is destroyed (almost completely). For the pristine case, there are two free-up-spin-electrons at one-edge and two down-spins at the other, and hence, a zero spin-moment. For both B- or N-ed passivated cases, free electrons of one spin will be destroyed leaving the other, and hence, a total spin-moment ≈ two. For both-ed passivated case, there will be no-free electrons, and hence, the spin-moment is zero.

### 3. 2.2. Systems with 1-line-PH-defect at single edge

From the **table 1**, it is clear that, only pristine and defect-edge passivated ribbons have non-zero $S_{pol}$, and this is because, in pristine and defect-edge passivated ribbons, zigzag edge of these ribbons is not passivated, and hence, there is a net $S_{pol}$. For the other two cases, system's zigzag edge is passivated, and hence, a zero $S_{pol}$ (as mentioned in with-out-defect ribbons). In accordance with the previous studies [9-11] reconstructed edge is always (in both bare and passivated cases) non-magnetic (because of its armchair edge nature).

### 3. 2.3. Systems with 1-line-PH-defect at both edges

As these systems have 1-line-PH-defect (arm-chair) at both the edges, one should expect that the systems should show zero $S_{pol}$ and indeed the expectation is correct.

### 3. 2.4. Systems with 2-line-PH-defect at one edge

As both edges are zigzag, as expected, we have non-zero $S_{pol}$ (**see table 1**) when a single-edge is passivated, because of the availability of the free-electrons of a particular spin, and, zero $S_{pol}$ when both edges are either passivated or pristine.

### 3. 2.5. Systems with 2-line-PH-defect at both edges

Again, as the both edges are zigzag, we have non-zero $S_{pol}$ when only one edge is passivated and zero $S_{pol}$ when both edges are pristine and passivated.

### *3.3. Electronic and magnetic properties:*

In DOS and pDOS plots, we have used the terms majority and minority-spins (depending on the value of $S_{pol}$) rather than the up and down-spins (as they are relative). If $S_{pol}$ (according to the above definition) is positive, then up-spin will be called as the majority spin and down-spin as minority and vice-versa. Whether a particular spin is majority or minority is indicated in the legends of the plots.

### 3.3.1. Systems without defect

A further inspection based on the wave-function analysis shows that (figures are not shown here), in bare ribbons, the majority-spin's valence-band-edge (VBE) is mainly composed of 2s and $2p_y$ orbitals of the B-edge atoms and the conduction-band-edge (CBE) is mainly composed of $2p_z$ and $2p_y$ orbitals of the N-edge. On the other hand, for the minority-spin channel, VBE is from the $2p_z$ orbitals of the N-edge and CBE is from the 2s and $2p_y$ orbitals of the B-edge atoms. As the system is $sp^2$ hybridized, $2p_z$ orbitals of the N-edge correspond to its lone-pair electrons and the 2s, $2p_y$ orbitals of the B and N-edge represents the hybridized-orbitals. So, the dangling bonds in these systems have a major contribution from the 2s, $2p_y$ orbitals and these states disappeared once the edges are passivated. Also, the higher reactivity of bare N-edge may help to react with H2 molecule more rigorously compared to B-edge (see section 3.1.1 of SI). For the case of both-edge passivated systems, there are no extra levels across the Fermi level and the systems are insulating with a band-gap similar to that of the minority-spin's band-gap in the bare and single-edge passivated systems. So, in ribbons without defect, systems are: 1) half-metallic when both the edges are pristine, 2) semi-conducting when one edge is passivated and 3) insulating when both edges are passivated. Thus, using passivation as a tool, we can tune these systems *majority-spin from metallic to semi-conducting to insulating*, while keeping the minority-spin always as insulating!

It is important to mention that, our study shows that B-edge passivation of a 10-zBNNR makes the system semi-conducting and this result contradicts with the previous studies[2, 12] [1]which have shown that B-edge passivation leads to half-metallicity. The reason is, all of these studies have been performed on 8-zBNNR and we also found that, for 8-zBNNR, B-ed-pa leads to half-metallicity (results not included here) and indeed our results matches exactly with the results of Lai. et. al [2]  (mainly because of the consideration of similar kind of exchange correlation and basis sets). The fluctuation of the half-metallic nature of B-ed-pa zBNNRs with ribbon widths (above the width of 8-zBNNR) has been observed recently by our group and will be discussed elsewhere [13].

### 3.3.6. Effect of Spin-configuration

In the perfect-edge systems we observed that, the boron edge states (represented in dashed magenta color) which are broad and spin-polarized in the (UU, DD) [or (DD, UU)] configuration changed to narrower and non-spin-polarized in the (UD, UU) configuration. This can be explained roughly as follows (please see [5] for further information): We know that the area under the peak of the DOS plot will give the number of energy-states. So, area under the B-edge states will give the number of such edge-states in that energy-range and they should be constant irrespective of the spin-configuration. Now, as we are changing the spin-configuration from (UU, DD) to (UD, UU) we are bringing spin-symmetry at the B-edge which urges equal contribution of the DOS for both- spins. As the number of edge-states can't change with spin-configuration and as both-spins should have equal contribution, the broad peak has to narrow down without changing the area under the curve and this is what we observed.

Except for slight narrowing in the peak-width, defect-edge systems show a different behavior compared to the perfect-edge systems. Unlike the perfect-edge systems, these systems show the spin-polarization even after a change in the spin-configuration from (UU, DD) to (UD, UU). Although we initially thought that there might be a difference in the distance between the two boron-edge atoms in perfect and defect edges, which can lead to different spin-spin interactions. But, this thought was over-thrown as the difference is just 0.001 Å.

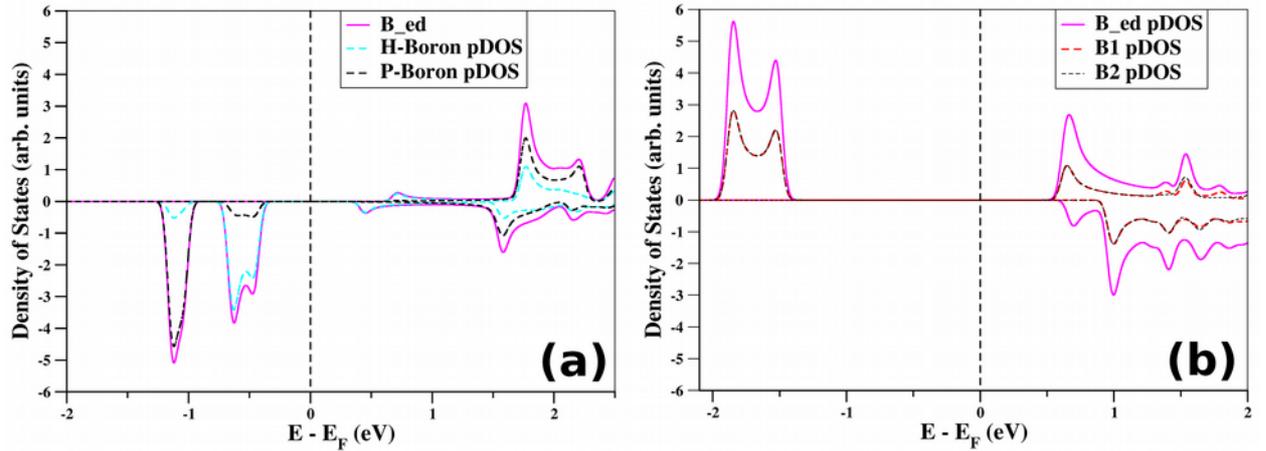

**Figure S4**: pDOS plots of each boron edge atom in the systems (a) B-2LD-NR-NH and (b) HBN-1LD-NR-B. H-Boron [P-Boron] means a boron-atom from a Heptagon [pentagon] ring.

In order to understand the cause for this difference in the perfect and defect edge systems, we have plotted the pDOS for each edge-atom for the systems B-2LD-NR-NH and HBN-1LD-NR-B, as a representative candidate of defect and perfect edge systems in figure S4a and S4b, respectively. Reason can be clearly understood by comparing both these figures. Figure S4a shows, for the defect-edge, each peak below the Fermi-level has a contribution mainly from a particular edge atom, and hence, when the spin on one atom is changed only that peak which has the contribution from that atom will move from one-spin state to the other. On the other hand, for the perfect edge (see figure S4b), all the peaks have equal contribution from both the edge atoms, and hence, changing the spin on any edge-atom will have an effect on all the peaks. Finally, the edge atoms are contributing to different peaks in defect-edge systems because the edge atoms themselves are different (one boron atom is from a heptagon ring and the other is from a pentagon ring), whereas, in the perfect edge system both the edge atoms are same (as they are arising from two hexagon rings). This gives us a hint that spin-spin interaction is negligible between the edge atoms in a defect edge system than in a perfect edge system.